\documentclass[ ]{jfm}
\usepackage{subcaption}

 \usepackage{setspace}
\usepackage{soul}
\usepackage{xcolor}

\usepackage{amsmath}
\usepackage{float}
\usepackage{graphicx}
\usepackage{newtxtext}
\usepackage{newtxmath}
\usepackage{natbib}
\usepackage{hyperref}
\hypersetup{
    colorlinks = true,
    urlcolor   = blue,
    citecolor  = black,
}

\newcommand{\RomanNumeralCaps}[1]
\linenumbers
       
\title{The effect of side walls on the stability of falling films}
\author{Hammam Mohamed\aff{1},
  J{\"o}rn Sesterhenn\aff{1}    \corresp{\email{Joern.Sesterhenn@uni-bayreuth.de}}
 \and Luca Biancofiore\aff{2}}
\affiliation{\aff{1}Lehrstuhl f{\"u}r Technische Mechanik und Str{\"o}mungsmechanik,  Universit{\"a}t Bayreuth, Bayreuth, Germany.
\aff{2} Department of Mechanical Engineering, Bilkent University, Ankara, Turkey. }

\begin{document}
\maketitle

\begin{abstract}
We study the influence of side walls on the stability of falling liquid films. We combine a temporal biglobal stability analysis based on the linearized Navier-Stokes equations with experiments measuring the spatial growth rate of sinusoidal waves flowing downstream an inclined channel.  Very good agreement was found when comparing the theoretical and experimental results. Strong lateral confinement of the channel stabilises the flow.  In the wavenumber-Reynolds number space, the instability region experiences a fragmentation due to selective damping of moderate wavenumbers.  For this range of parameters, the three dimensional confined problem shows several prominent stability modes which are classified into two categories, the well known Kapitza hydrodynamic instability mode (\textit{H-mode}), and a new unstable mode, we refer to it as wall-mode (\textit{W-mode}).  The two mode types are stabilised differently, where the H-modes are stabilized at small wavenumbers, while the W-modes experience stabilization at high wavenumbers, and at sufficiently small channel widths, only the W-mode is observed. { The reason behind the unique H-modes stabilisation is that they become analogous to \textit{waveguide} modes, which can not propagate below a certain cut-off wavenumber.}  The spatial structure of the eigenmodes experiences significant restructuring at wavenumbers smaller than the most damped wavenumber. The mode switching preserves the spatial symmetry of the unstable mode.  
\end{abstract}


\section{Introduction}

 {Falling liquid films are thin viscous liquid flows driven down an inclined plate by gravity. For several decades, they have been under extensive investigation due to their important role in various mechanical and chemical applications, such as complex flow coating \citep{weinstein2004coating} and  cooling of mechanical and electronic systems \citep{hu2014numerical}. For instance, the  wavy nature of falling films organized into localised structures or solitary waves \citep{miyara2000numerical} were found to increase the heat/mass transfer rate by $10-100\%$ compared to flat films \citep{goren1968mass}. Therefore, understanding the mechanisms leading to the complex dynamics of falling liquid films are crucial to the development of various industrial applications.  Studying the linear stability is considered the first  step in  understanding the disordered spatio-temporal state that characterizes falling liquid films. Investigating the onset of the instability and the domains in which the flow is stable or unstable is crucial to understand this kind of flows.}

 {Mostly, the stability of falling liquid films with respect to infinitesimal perturbations is studied using the \textit{modal} approach, in which the eigenmodes of an operator describing the evolution of the perturbations are examined separately. This approach relies on decomposing the flow quantities ${{\bf q}}$ into  a steady part $\bar{{\bf q}}$ (usually  called \textit{base state}) and an unsteady part $\tilde{{\bf q}}$ (perturbations). Traditionally, the base state of a falling film is found using the \textit{parallel flow} approximation, where the wall-normal velocity component and the spanwise derivatives are assumed to be negligible. With this approximation, the base state forms a semi-parabolic velocity profile varying in the wall-normal direction, while it remains constant in the other directions. This base state was first obtained by \cite{nusselt1916oberflachenkondensation}, and is referred to as \textit{Nusselt film solution}.}

 {In the case of laterally confined falling films, the base state does not only change in the direction normal to the bottom wall, but also in the spanwise direction. \cite{scholle2001exact} followed by \cite{haas2011side} used analytical and experimental methods to investigate the effect of side walls on the base state of falling films. They found two competing effects in the vicinity of the wall; (i) the reduction of the base  flow velocity due to the no-slip boundary condition; (ii)   an increase in the surface velocity near the side walls due to the increased film thickness caused by the capillary elevation. However, the latter effect can be ignored if the film thickness is sufficiently large.  }

 {The linear stability analysis based on the Nusselt base state is a \textit{local stability analysis}, since the flow stability at a single point in the downstream is sought. The stability analysis in which the base flow varies in two spatial directions, as in laterally confined falling films, is named \textit{biglobal stability analysis}. Finally, when the base state is not homogeneous in any of the considered spatial directions, the stability analysis is referred to as \textit{global stability analysis}, for example see \cite{albert2014global}. A quantitative comparison between different stability modal approaches can be found in the work of \cite{juniper2014modal}.}

The pioneering works of \cite{kapitza1948wave,kapitza1949experimental} were first to describe the development of "long" wavelength deformations on the surface of the film. They identified these deformations as hydrodynamic instability, and therefore called it \textit{H-mode} instability.  The Orr-Sommerfeld approach has been followed in much of the work devoted to studying the local stability of falling films with respect to the hydrodynamic instability.  \cite{benjamin1957wave} and \cite{yih1963stability} were first to follow this approach as they performed a temporal stability analysis in which the wave velocity and amplification factor for a perturbation with a given wavelength is sought. They identified the onset of the H-mode for small wavenumbers  in terms of the critical Reynolds number and inclination angle $\beta$ as $Re_c = (5/6) \cot(\beta)$, {where this condition is only valid when the Reynolds number is based on the average liquid velocity}. On the other hand, \cite{krantz1973spatial} performed a spatial stability analysis where the spatial amplification and the wavelength of a disturbance with a prescribed frequency is sought. A more simplified approach was formulated by \cite{benney1966long}, where he used long wave expansion to obtain a temporal evolution equation (\textit{Benny's} equation) that describes the interface development for small wavenumber perturbations. The evolution equation can be utilized to obtain a dispersion relation of the temporal amplification rate in the regions of small wavenumbers and weak inertia. Various physical effects were considered while studying the local stability of falling films such as heating and phase change \citep{mohamed2020linear}, shearing gas \citep{lavalle2019suppression}, and flexible inclined surface \citep{alexander2020stability}, etc. For the wide range of mathematical approaches and physical effects considered while studying the interface evolution and local stability of falling films, refer to the reviews by \cite{oron1997long}, \cite{craster2009dynamics}, and \cite{kalliadasis2011falling}.

{ With regards to the linear stability of laterally confined falling films (biglobal stability), \cite{vlachogiannis2010effect} and \cite{georgantaki2011measurements} experimentally studied long-wave perturbations in narrow channels. The earlier showed that the critical  Reynolds number ($Re_c$) increases significantly as the channel width decreases, while it converges to the two dimensional theoretical value as channel width increases. The latter, \cite{georgantaki2011measurements}, confirmed the aforementioned observation, and additionally showed that surface tension, opposed to classical stability, has a drastic effect on the longwave instability threshold. They showed that the  deviation in the critical Reynolds number due to spanwise confinement is strongly enhanced as the ratio between capillary and viscous forces increases.} \cite{pollak2011side} investigated the influence of the contact angle at the side walls on the onset of the stability. Experiments showed a significant change in the neutral stability curve between two different contact angles, but with no structural modification on the neutral curves. Most recently, \cite{kogel2020stability} performed experiments for a wide range of forcing frequencies and Reynolds numbers. They showed that the side walls can have a dramatic effect on the shape of the neutral curve by causing a fragmentation due a selective damping of the instability at moderate frequency range. They concluded that the damping effect is independent of the inclination angle and only depends on the channel width.

 {While the effect of spanwise confinement on the stability of falling films is not properly investigated from the theoretical perspective, there are several examples in the literature for other flow configurations. \cite{tatsumi1990stability} formulated two Orr-Sommerfeld like equations to study the the stability of laminar flow in a rectangular duct. Their stability analysis showed that the critical Reynolds number increases monotonically with decreasing the aspect ratio of the duct, thus a stabilising effect is observed.  \cite{theofilis2004viscous} showed similar results solving the full linearized Navier-Stokes equations. More recently, \cite{lefauve2018structure} and \cite{ducimetiere2021effects} studied the effect of spanwise confinement on stratified shear instabilities showing that the lateral confinement has a stabilising effect. }

 {	{We examine the effect of side walls on the linear stability of falling liquid films by combining theoretical modeling and experimental techniques. This  approach to the problem introduces the missing theoretical background for the spanwise confinement effect reported in the literature, which is limited to experiments only. Additionally, the theoretical stability model offers the opportunity to explore a wide range of flow parameters, which was not possible due to the experimental techniques restrictions.} This monograph is organized as follows, In section \ref{sec:Theore_form}, we discuss the non-dimensional governing equations alongside the dimensionless parameters. The base state and the linear stability analysis are also presented in the same section. The experimental set up and the linear stability measurement technique are presented in section \ref{sec:exp_Setup}. Our results consisting of validating the numerical stability model with the experiments followed by extensive theoretical investigation of different aspects of the problem are presented in section \ref{sec:results}. Finally, we present our concluding remarks and potential direction of investigations in section \ref{sec:conclusion}.  
 	
\section{Theoretical formulation}\label{sec:Theore_form}
 {The theoretical formulation of the problem is presented in this section. We start with listing the non-dimensional governing equations, and the associated walls and the free surface boundary conditions, in section \ref{goveq}. Afterwards, we present the base state solution (section \ref{base}) followed by the methodology to linear stability analysis (section \ref{linear_Stability}). Finally, the numerical approach to solve the problem is presented in section \ref{numerical_method}. }
\subsection{Non-dimensional governing equations:} \label{goveq}

\begin{figure}
	\centerline{\includegraphics{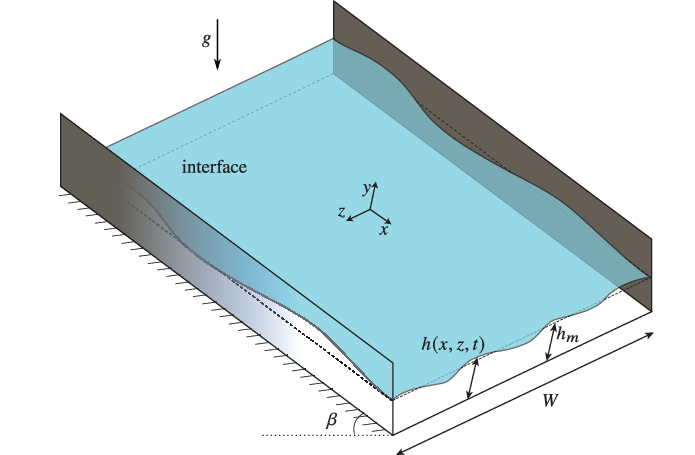}}
	\caption{Schematic diagram of liquid film falling down an inclined channel. $h(x,z,t)$ is the local film thickness, and $h_m$ is the mean film thickness.}
	\label{fig:fig1}
\end{figure}

We consider a three-dimensional liquid film running down a tilted channel due to gravity $g$ (figure \ref{fig:fig1}). The channel has a width $W$ and forms an angle $\beta$ with the horizontal. The density $\rho$ and dynamic viscosity $\mu$ are constant, while the kinematic viscosity $\nu$ is given by $\nu = \mu/\rho$. We use the Cartesian coordinates $(x,y,z)$, where $x$ corresponds to the streamwise direction, $y$ is bottom wall normal increasing into the liquid, and $z$ is in the spanwise direction. The left and right walls of the channel are located at $z = 0$ and $z = W$, respectively. The following length, time, velocity and pressure scales are introduced in order to non-dimensionalize the system:
 \begin{center}
	\begin{tabular}{ccc}
	$ \quad	h\rightarrow h_m \ h^*,\quad$ & $\quad(x,y,z) \rightarrow h_m \ (x^*,y^*,z^*),\qquad$ & $ \quad W\rightarrow h_m \ W^*,\quad$\\ \\
	$\quad t  \rightarrow h_m^2/ \nu \ t^*,\quad$ & $\quad (u,v,w)  \rightarrow \nu /h_m \ (u^*,v^*,w^*),\quad$   &  $ \quad p  \rightarrow  \rho \nu^2 / h_m^2 \ p^*,\quad$\\ \\
\end{tabular}
 \end{center}
{where $h_m$ is the mean film thickness}. By using these scales and {dropping the stars for simplicity}, we obtain the dimensionless governing equations, continuity and Navier-Stokes, as follows:

\begin{subeqnarray}
	& \partial_x {u} + \partial_y {v} + \partial_z {w} = 0, \\
	&  \partial_t u +\boldsymbol{u} \cdot \nabla {u} = - \partial_x p + \nabla^2 {u} + G \sin(\beta), \\
	&  \partial_t v +\boldsymbol{u} \cdot \nabla {v} = - \partial_y p + \nabla^2 {v} + G \cos(\beta), \\
	&  \partial_t w +\boldsymbol{u} \cdot \nabla {w} = - \partial_z p + \nabla^2 {w},
\end{subeqnarray}\label{governing}
where $G = g h_m^3/\nu^2 $ is the ratio of inertial to viscous forces. The Reynolds number is a function of the flow rate $\dot{q}$ and channel width $W$: 
\begin{equation}
	Re = \frac{\dot{q}}{2 \nu W}.
\end{equation} 	\label{Reynolds}
  
It is important to clarify that $G \sin(\beta)$ does not represent the Reynolds number as in the  case of the two dimensional problem ($W \rightarrow \infty$). The quantity $G \sin(\beta)$ is always larger than the Reynolds number because the velocity is slowed down in the vicinity of the wall. $Re$ approaches $G \sin(\beta)$ as the channel width goes to infinity. The dimensionless boundary conditions at the bottom $(y=0)$ and side walls $(z=0,W)$ read:

\begin{subeqnarray}
	&\boldsymbol{u}(x,y=0,z) = 0,\\
	& \boldsymbol{u}(x,y,z = (0, W)) = 0.
\end{subeqnarray}\label{wallcond}
 Moreover, the kinematic and dynamic couplings at the free surface lead to the interface boundary conditions at $y=h(x,z,t)$ which are written as: 
 
\begin{subeqnarray}
	&v = \partial_t h + u  \partial_x  h + w \partial_z,\\
	&p =  \frac{2}{n^2} \Big[   (\partial_x h)^2 \partial_x u + (\partial_z h)^2 \partial_z w + \partial_x h \partial_z h (\partial_z u + \partial_x w) \nonumber  \\
	&\mbox{} -  \partial_x h (\partial_y u + \partial_x v)- \partial_z h (\partial_z v + \partial_y w) + \partial_y v\Big] \nonumber\\
	& - \frac{1}{n^3} 3S \Big[ \partial_{xx} h(1+(\partial_z h)^2) + \partial_{zz} \small(1+(\partial_x h)^2) - 2\partial_x h \partial_z h \partial_{xz} h \Big] ,\\
	&0 =  \frac{1}{n} \Big[   2\partial_x h (\partial_y v -  \partial_x u) + (1-(\partial_x h)^2)(\partial_y u + \partial_x v) - \nonumber \\ 
	&\partial_z h (\partial_z u + \partial_x w) - \partial_x h \partial_z h(\partial_z v + \partial_y w),\\
	&0 =  \frac{1}{n} \Big[   2\partial_z h (\partial_y v -  \partial_z w) + (1-(\partial_z h)^2)(\partial_y w + \partial_z v) \nonumber \\
	&- \partial_x h (\partial_z u + \partial_x w) - \partial_x h \partial_z h(\partial_y u + \partial_x v) ,
\end{subeqnarray}\label{interfacecon}
where $n = (1+(\partial_x h)^2 + (\partial_z h)^2)^{1/2}$. The kinematic interface boundary condition ({2.4a}) governs the relationship between the film thickness and the normal velocity component, while the equilibrium between the pressure and surface tension at the interface is governed by the normal and tangential stress boundary conditions ({2.4b-2.4d}), where the non-dimensional surface tension is written as $S = {\sigma h_m}/{3 \rho \nu^2}$. For the complete derivation of the boundary conditions see \cite{kalliadasis2011falling}. 

{Finally, we need to address the wetting effects which are also tied to the contact line behaviour at the side walls where solid, liquid and ambient meet. Formulating an accurate description of the  intertwined issues of meniscus and contact line dynamics is a complex task, and has received a considerable amount of attention in the literature \citep{snoeijer2013moving}. For a thin liquid film, wetting effects result in an increase in the local film thickness at the side walls. This increase is defined as $\Delta h = \ell_c \sqrt{2 (1- \sin(\theta))}$ where $\ell_c$ is the general capillary length ($\ell_c  = \sqrt{\sigma/\rho g \cos(\beta)}$). In the scope of this work, the Bond number ($\mbox{Bo} = W^2/\ell_c^2$), which measures the magnitude of gravitational forces to surface tension for interface movement, is very high since the capillary length is much smaller than the channel width ($\ell_c<<W$). Therefore, wetting effects are not dominant and are limited to a thin region near the side walls, and thus, neglected. Consequently, we implement a simple free-end boundary condition  where the contact line can freely slip at the side walls with a contact angle usually chosen as $\pi/2$:}
\begin{equation}
\partial_z {h}(y=h,z=(0,W)) = \pm \cot(\theta) = 0. 
\end{equation}\label{contactbc}
{For a follow-up discussion on wetting effects and contact line dynamics, the reader is referred to section (\ref{contact}). Additionally, wetting effects can cause an overshoot in the velocity in the vicinity of the side walls when the increase in the film thickness $\Delta h$ exceeds the mean film thickness $h_m$ as shown by \cite{scholle2001exact}. However, this condition is never met in our work and therefore, this effect is outside the scope our analysis.}


\subsection{Base flow} \label{base}
{The effect of side walls on the base flow with an undisturbed free surface is examined in this section. When the spanwise direction is unbounded ($W \rightarrow \infty$), the system presented in equations (\ref{governing}) alongside the walls and interface boundary conditions ({2.3-2.5}) has a one-dimensional base flow solution as a semiparabolic velocity profile known as the \textit{Nusselt film solution} \citep{nusselt1916oberflachenkondensation}. When the domain is confined in the spanwise direction, the base flow solution is altered because of (i) the no-slip boundary condition on the side walls which causes the velocity to decrease in the vicinity of the side walls, and (ii) the wetting effects causing a capillary elevations at the side walls. Regarding the latter effect, we neglect it by setting the contact angle to $\pi/2$  based on the  discussion in section \ref{goveq}.}

Next, by assuming the bottom wall-normal velocity ($v$) and the streamwise spatial derivatives ($\partial_x$) to be zero, the base flow of the three-dimensional problem $(3D)$ can be found numerically by solving the following system of equations:

\begin{subeqnarray}
	&\partial_{yy} \bar{U} + \partial_{zz}\bar{U} = G \sin(\beta),  \\ 
	&\partial_z \bar{P} = 0. 
\end{subeqnarray}
With the walls and interface boundary conditions:

\begin{subeqnarray}
	& \bar{U}(y=0,z) = 0,  \qquad &\mbox{ } \\
	& \bar{U}(y,z=(0,W)) = 0,	       \qquad &\mbox{and} \qquad \partial_z \bar{h}(z= (0, W)) = 0, \\
	& \partial_y \bar{U}(y=h,z) = 0, \qquad &\mbox{and} \qquad \bar{P}(y=h,z) = 0. 
\end{subeqnarray}

\begin{figure}
	\centerline{\includegraphics{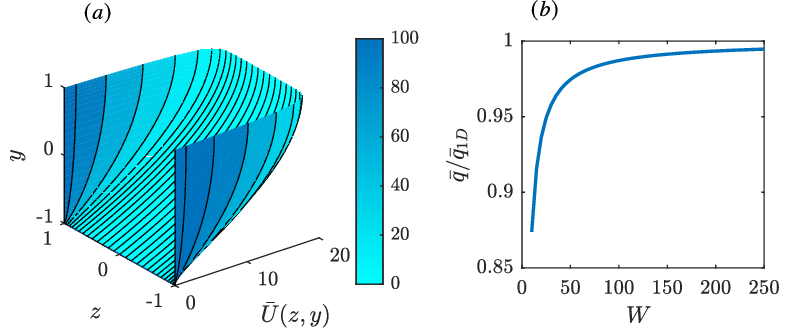}}
		\caption{Base state velocity for a channel with $W=20$ for $G \sin(\beta) = 40$. (b) Normalized local flow rate per unit width $\bar{q}$.}
	\label{fig:basestate}
\end{figure}
%


 {Figure \ref{fig:basestate}(a) shows the base state velocity for a confined channel with $W=20$. The color scale presents the percentage drop in the velocity profile $\bar{U}(z,y)$ from the $1D$ Nusselt velocity profile $\bar{U}_{1D}(y)$ along the $y$-direction. Clearly, the base state is only affected in a small region near the side walls, in which the velocity is slowed down, while it retains the Nusselt film solution in most of the channel width. We approximate this region to be in the order of the mean film thickness $h_m$. This is expected since the side walls effect on the velocity is mainly controlled by the ratio between the film thickness and the channel width which is typically small, for example, for a relatively narrow channel of $W=20$, this ratio equals $0.05$. }
 
{ Moreover, the local flow rate per unit width ($\bar{q}$) along the channel width $W$ is shown in figure \ref{fig:basestate}(b). The values are normalised with the flow rate per unit width for an infinitely wide channel ($\bar{q}_{1D}$), which is in fact equals to $G \sin(\beta)/3$. The local flow rate does not experience a significant drop even for relatively narrow channels. For example, the flow rate is only  decreased by $5\%$ compared to the 1D flow rate for a channel width of $W=25$. The local flow rate asymptotically reaches $\bar{q}_{1D}$ as the channel width increases. This is again because the effect of side walls on the flow is restricted to a thin region in the vicinity of the walls, while outside this region, the flow obeys the 2D Nusselt flow.\\
}

\subsection{Linear stability analysis} \label{linear_Stability}
 The flow field variables $\boldsymbol{q}(x,y,z,t) =(u,v,w,p$) in addition to the interface variable $h(x,z,t)$ are expanded as a sum of the base flow  and an infinitesimal perturbation as:

\begin{subeqnarray}
	&\boldsymbol{q}(x,y,z,t) = \bar{\boldsymbol{q}}(y,z,t) + \epsilon \tilde{\boldsymbol{q}}(x,y,z,t),\\
	&	h(x,z,t) = 1 + \epsilon \tilde{h}(x,z,t).
\end{subeqnarray}  \label{expansions}

Expansions in {2.8} are then substituted in the governing equations (\ref{governing}) and boundary conditions ({2.3--2.5}), which are then linearized for $\epsilon <<1$ leading to the linearized perturbation equations at $O (\epsilon)$ :

\begin{subeqnarray}
	& \partial_x \tilde{u} + \partial_y \tilde{v} + \partial_z \tilde{w} = 0, \\
	& \partial_t \tilde{u} + \bar{U} \partial_x \tilde{u} + \partial_y \bar{U} \tilde{v} + \partial_z \bar{U} \tilde{w} + \partial_x \tilde{p} - \nabla^2 \tilde{u} = 0, \\
	& \partial_t \tilde{v} + \bar{U} \partial_x \tilde{v} + \partial_y \tilde{p} - \nabla^2 \tilde{v} = 0,\\
	& \partial_t \tilde{w} + \bar{U} \partial_x \tilde{w} + \partial_z \tilde{p} - \nabla^2 \tilde{w} = 0,
\end{subeqnarray}
with the bottom and side wall boundary conditions:

\begin{subeqnarray}
	&\tilde{\boldsymbol{u} }(x,y=0,z) = 0, \\
	&\tilde{\boldsymbol{u}}(x,y,z=(0, W)) = 0.
\end{subeqnarray}
For the interface boundary conditions at $y= 1+\tilde{h}$, we utilize a Taylor expansion and expand the equations around the undeformed interface $y=1$, for example the variable $X$ is expanded as follows $X|_h = X(1) + \tilde{x}|_1 + DX(1) \tilde{h}$:

\begin{subeqnarray}
	&\tilde{v} - \partial_t \tilde{h} - \bar{U} \partial_x \tilde{h} = 0,\\
	&\tilde{p} + \partial_y \bar{P} - 2 \partial_y \tilde{v} + 3S \nabla^2_{xz} \tilde{h} = 0, \\
	& - \partial_{yy} \bar{U} \tilde{h} - \partial_y \tilde{u} - \partial_x \tilde{v} + \partial_z \tilde{h} \partial_z \bar{U} = 0, \\
	& \partial_y \tilde{w} + \partial_z \tilde{v} - \partial_z \bar{U} \partial_x \tilde{h} =0.
\end{subeqnarray}

Afterwards, the pressure perturbation boundary conditions are needed at the walls in order to close the problem. We  use the solution proposed by \cite{theofilis2004viscous}, which is to introduce the compatibility condition at the walls derived from the Navier-Stokes equations as follows: 

\begin{subeqnarray}
	&  \partial_y \tilde{p}  = \nabla^2 \tilde{v} - \bar{U} \partial_x \tilde{v},\\
	&  \partial_z \tilde{p}  = \nabla^2 \tilde{w} - \bar{U} \partial_x \tilde{w}.
\end{subeqnarray}

Finally, we introduce Fourier modes to expand the  perturbation variables in $x$ and $t$ as: 

\begin{subeqnarray}
	&  	\tilde{\boldsymbol{q}}(x,y,z,t) = \boldsymbol{Q}(z,y) \mbox{ exp}(ikx - i\omega t),\\
	& \tilde{h}(x,z,t) = \eta(z) \mbox{exp}(ikx - i\omega t),
	\label{normal_modes}
\end{subeqnarray}
where $k$ is the wavenumber, and $\omega$ is the angular frequency. For \textit{temporal stability analysis}, $k$ is considered to be real, and we solve for the complex eigenvalue $\omega$. The \textit{temporal growth rate } $\omega_i$ is the imaginary part of $\omega$. If $\omega_i >0$ the perturbation grows in time and the base flow is unstable. Moreover, the \textit{phase velocity} is denoted as $c_r = \omega_r/k$. For \textit{spatial stability analysis}, $\omega$ is assumed real, and $k$ is complex. Similarly, the flow is unstable if the \textit{spatial growth rate} $\gamma_i$ is larger than zero. {Equation (\ref{normal_modes}) may describe waveguide modes \citep{lighthill2001waves}, where the wave disturbances are allowed to travel in the longitudinal direction while being confined in the spanwise direction. They can be interpreted physically as plane modes ($W \rightarrow \infty$) bouncing back and fourth between the side walls \citep{elmore1985physics}. This concept is revisited in the next sections as it leads to interesting results.}

\subsection{Numerical method} \label{numerical_method}
We discretize the domain using two-dimensional Chebyshev differentiation matrices \citep{trefethen2000spectral}. The boundary conditions were implemented by replacing the corresponding rows in the matrices with the discretized boundary conditions. 
We prefer a temporal stability analysis since it requires less storage and computational power than what is required by the spatial stability problem. We base our results mainly on the temporal growth rate $\omega_i$, and utilize Gaster transformation \citep{gaster1962note} that allows us to relate the temporal and spatial growth rates as:
\begin{align}
	\gamma  = \omega_i/ \frac{\partial \omega_r}{\partial k} 
	\label{gaster}.
\end{align}

\cite{brevdo1999linear} showed that this transformation is applicable to falling liquid films since the spatial and temporal growth rates are small. Finally, the eigenvalue problem was solved using the QZ algorithm \citep{moler1973algorithm}.  

\begin{figure}
	\centerline{\includegraphics{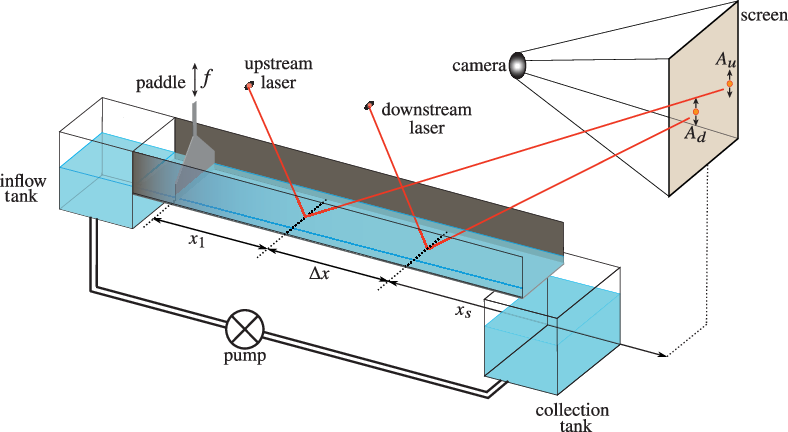}}
	\caption{Sketch of the experimental set-up used to measure linear stability.}
	\label{fig:expsketch}
\end{figure}

\section {Experimental setup}\label{sec:exp_Setup}
We use the very experimental rig as  \citep {pollak2011side,schorner2016stability,kogel2020stability}. Here, we briefly present the experimental apparatus and measurement technique. For more details, please refer to the aforementioned works.
\subsection{Experimental apparatus}
The experimental set-up we use in this work is sketched in figure \ref{fig:expsketch}.  The set-up consists of a channel with an aluminum bottom and Plexiglas side walls. The width of the channel is $170$ mm, but two movable side walls can be placed within the channel to achieve smaller channel widths. The whole channel is tilted with an inclination angle $\beta$ and mounted onto a damped table in order to reduce the effect of vibrations. A collection tank is located at the end of the channel in which the liquid is maintained at a certain temperature $T$. An eccentric pump is used to convey the liquid from the collection tank to the inflow tank located at the inlet of the channel. The inflow tank allows the liquid to rest and to smoothly overflow into the channel. Periodic waves are introduced in the flow using a paddle that is permanently dipped in the liquid film. The paddle oscillates at an adjustable frequency $f_p$ and amplitude $A_p$, causing sinusoidal waves evolving downstream the channel with the same frequency and amplitude. {The disturbances generated by the paddle are instantaneously convected downstream the channel, and do not disturb the flow in the inflow tank.} 
The working liquid used in this experiment is \textit{Elbesil 140} which is a mixture of silicon oils, \textit{Elbesil 50} and \textit{Elbesil 200}. The physical properties of \textit{Elbesil 140} at temperature $24^{\circ}$ C are listed in table \ref{tab:fluid_properties}.

\begin{table}
	\begin{center}
		\def~{\hphantom{0}}
		\begin{tabular}{lcc}
			Physical property  & Notation  &   value \\[3pt]
			Density   & $\rho$ & $962.7 \pm 0.4 \ \  (\mbox{kg} \ \ \mbox{m}^{-3}  $) \\
			Kinematic viscosity   & $\nu$ & $142 \pm 0.5 \ \  (\mbox{mm}^2 \ \ \mbox{s}^{-1}$) \\
			Surface tension  & $\sigma$ & $20.25 \pm 0.11 \ \  (\mbox{mN} \ \ \mbox{m}^{-1})$\\
		    Contact angle & $\theta$ & $17.25 \pm 7 \ \  (\mbox{Degree})$\\
		\end{tabular}
		\caption{Physical properties of \textit{Elbesil 140} at working temperature $T = 24^{\circ} \mbox{C}$.}
		\label{tab:fluid_properties}
	\end{center}
\end{table}

\subsection{Linear stability measurement} \label{Linear stability measurement}
The linear stability of the flow was investigated through measuring the spatial growth rate of the excited waves $\gamma$. Two laser beams are reflected at the film's surface at two locations, upstream and downstream. The two laser reflections are projected onto a screen on which they oscillate at the same frequency of the sinusoidal waves generated by the paddle. The oscillations of the laser spots on the screen can can be written as:

\begin{subeqnarray}
	&A_u(t) = A_u^0 \sin(2\pi f t),   \\ 
	&A_d(t) = A_d^0 \sin(2\pi f t), 
\end{subeqnarray}
where $A^0$ is the amplitude of the signal, and the subscripts $u$ and $d$ correspond to upstream and downstream locations, respectively. Simply, the oscillation amplitudes $A_u^0$ and $A_d^0$ are proportional to the oscillation amplitude of the liquid interface slope, which is also proportional to the waves amplitudes. This is true as long as the waves are linear and sinusoidal, for detailed discussion please see the work of \cite{pollak2011side}. A CCD camera with a 100 frames per second is used to capture the screen for 10 seconds, resulting in 1000 images each measurement.  Figures \ref{fig:f10_osc}(a,b) show  the oscillation of the upstream and downstream laser spots, for a paddle frequency $f_p$ equals $10$ Hz with an amplitude $A_p=$  $0.6$ mm. The oscillations of the laser spots were obtained by applying a Gaussian filter on the images and tracing the centres of the laser spots. The mean film thickness $h_m$ was measured to be $9$ mm, and, thus, the excited waves amplitude is $6.6 \%$ of the film thickness. The excited waves amplitude was chosen large enough to be adequately examined, but small enough to be compatible with the linear stability theory.
\begin{figure}
	\centerline{\includegraphics{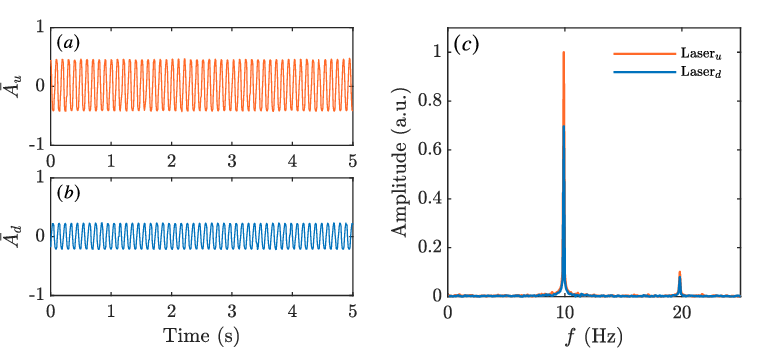}}
	\caption{Normalized upstream and downstream signals and their corresponding discrete Fourier transform for forcing frequency $f = 10$ Hz and Amplitude $A_p = 0.6$ mm. The peaks are located at $f=9.91$ Hz.}
	\label{fig:f10_osc}
\end{figure}
Subsequently, the amplitudes $A_u^0$ and $A_d^0$ were found by performing discrete Fourier analysis of the laser spots oscillations traced from the images. Figure \ref{fig:f10_osc}(c) shows  the Fourier spectra for the same paddle frequency and amplitude ($f_p=10, A_p=0.6$). The spectra show a dominant peak indicating that the signal is monochromatic with a frequency very close to the excitation frequency (maximum difference between the excitation frequency and response frequency is $0.1$ Hz). 

The downstream amplitude was geometrically corrected since it covers a smaller distance from the reflection point to the screen as follows: 
\begin{equation}
	A^0_{d,\mbox{corr}} = 	A^0_d \ \Big(1 + \frac{\Delta x}{x_S}\Big),
\end{equation}
where, $\Delta x = 400$ mm  is the distance between the two reflection points, and $x_S = 2.966 $m is the distance from the downstream reflection point to the screen. Furthermore,  $x_1$ is the distance between the channel inlet and upstream laser spot, which we made sure to be sufficiently large to avoid inlet effects on the base flow. Finally, the linear growth rate of the excited waves is obtained as:

\begin{equation}
	\gamma = \frac{\ln(A^0_{d,\mbox{corr}} /A_u^0 )}{\Delta x}.
\end{equation}

\section{Results}\label{sec:results}
 We start with  confronting our numerical linear stability with experimental results in section \ref{results_valida}. Afterwards, in section \ref{results_th}, we utilize the numerical model to examine the effect of the spanwise confinement on several aspects of the problem, such as, (i) maximum linear growth rate, (ii) types of stability modes and their distinct behaviour, (iii) eigenmodes structure, and  (iv) eigenvalues spectra. 
\subsection{Comparison of experiments  and theory} \label{results_valida}
In this section, we compare the results obtained by our numerical stability model, our experiments, as well as the experiments conducted by \cite{kogel2020stability}. The comparison is based on the spatial growth rate $\gamma$, which is directly found from the experimental measurements as explained in section \ref{Linear stability measurement}. Gaster transformation is utilized to obtain the numerical spatial growth rate from the quantities $\omega$ and $k$ as shown in equation (\ref{gaster}). We also need to obtain the response frequency $f$ corresponding to each wavenumber $k$, which can be easily found as $f=\omega_r(k)/2\pi$. The experimental control parameters and their corresponding non-dimensional numbers used in the numerical model are listed in Table \ref{tab:exp_parameters}, {  in which the general capillary length is shown to be much smaller than the channel width ($\mbox{Bo} \approx 3850$), and therefore, the wetting effects can be neglected.}
\begin{table}
	\begin{center}
		\def~{\hphantom{0}}
		\begin{tabular}{lcc}
			Control parameter  & Notation  &   Range of variation  \\[3pt]
			Dimensional channel width (mm)   & $W$ & $129 \pm 0.5$ \\
			Inclination angel (Deg)  & $\beta$ & $10 \pm 0.1$\\
			Mean film height (mm)  & $h_m$ & $9.025\pm 0.02$ \\
			{Capillary length (mm)}    & {$\ell_c$} &  {$2.08 \pm 0.01$} \\
			Channel width/mean film thickness & $W/h_m$ & $14.3 \pm 0.087$  \\
			Flow parameter   &  $G $ & $357 \pm 4.9$ \\
			Surface tension    & $S$ & $3.19 \pm 0.047$ \\
		\end{tabular}
		\caption{Experiment parameters and non-dimensional numbers for the numerical stability model.}
		\label{tab:exp_parameters}
	\end{center}
\end{table}

\begin{figure}
	\centerline{\includegraphics{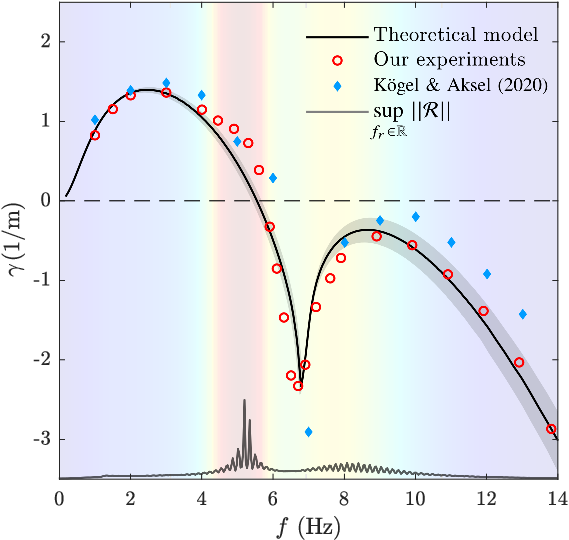}}
	\caption{Spatial growth rate obtained using our theoretical model, our experiments, and the experiments of \cite{kogel2020stability}. The color in background represents the amplitude of  $\sup_{f_r \in \mathbb{R}
		} ||\mathcal{R}||$ }
	\label{fig:expvsth}
\end{figure}

Figure \ref{fig:expvsth} shows the spatial growth rate in terms of the response frequency. In general, good agreement is observed between our numerical model and our experiments. Most importantly, the sharp cusp in the growth rate and its corresponding frequency are captured accurately. The minimal shift in the response frequency is due to the accuracy of the Fourier analysis which has an error of $\pm 0.1$ Hz. The grey band represents the upper and lower limits found by the numerical model. These limits were obtained based on the tolerances in the fluid properties and experiment control parameters (Tables \ref{tab:fluid_properties} and \ref{tab:exp_parameters}). For example, the upper limit was obtained by choosing the parameters limits that result in the maximum growth rate combining $(G_{max},S_{min},\beta_{max},W_{max})$, while the opposite was applied in finding the lower limit.

Furthermore, while our theoretical and experimental results match very well for low  and high  frequencies, an evident discrepancy is observed before and after the cusp around the frequencies $5.5$ and $7.5$ Hz. We offer a heuristic support for this discrepancy by plotting the supremum of the resolvent norm ($ \sup_{f_r \in \mathbb{R}
} ||\mathcal{R}||$), which indicates the maximum amplification over the excitation frequency $f_r$ \citep{trefethen1993hydrodynamic}:
\begin{equation}
	\sup_{f_r \in \mathbb{R}
	} ||\mathcal{R}|| = \sup_{f_r \in \mathbb{R}
}  || (-i f_r B+ A)^{-1} ||,
\end{equation}
where $A$ and $B$ are our matrices in the general eigenvalue problem. Since the streamwise spatial derivative is replaced by $ik$ based on the modal decomposition in equation (\ref{normal_modes}), this requires sweeping over the frequency $f_r$ to find $\sup_{f_r \in \mathbb{R}
} ||\mathcal{R}||$ at a certain wavenumber $k$. Finally, we find the corresponding response frequency for every wavenumber to be able to compare against the spatial growth rate, as shown in figure \ref{fig:expvsth}. It is obvious that the norm of the resolvent appears to be proportional to the discrepancy between the numerical and experimental results. The maximum value of $ \sup_{f_r \in \mathbb{R}
} ||\mathcal{R}||$ is located before the cusp at which the maximum discrepancy exists ($f \approx 5.5$ Hz), while the lower peak matches the smaller discrepancy region after the cusp ($f \approx 7.5$ Hz). Therefore, the discrepancy between the theoretical and experimental growth rate appears to be maximum where $\sup_{f_r \in \mathbb{R}
}||\mathcal{R}||$ is large. While the experimental growth rate accounts for the maximum amplification of the disturbances, the numerical growth rate is taken directly from the most unstable stability mode. 
\subsection{Theoretical results}\label{results_th}
We now study the effect of spanwise confinement on the linear stability from different perspectives. The numerical model allows exploring the parameters space, which was not possible using experimental means. To begin with, it is useful to recall the results of the well-studied one-dimensional ($1D$) local stability problem (where the spanwise variation is not taken into account) in order to accurately assess the influence of the side walls. The $1D$ problem results in only one prominent stability mode, which is responsible for the $1D$ hydrodynamic instability. For this reason, is it called {\it H-mode}. \footnote{We refer to $1D$ H-mode for distinction as $1D$-mode in this paper and preserve the name H-mode for the $2D$ equivalent. } We found that our problem results in several stability modes which behave differently depending on the degree of the spanwise confinement. In the following, we first focus on the maximum temporal growth rate regardless of which stability mode is dominant. Afterwards, we quantify the most important stability modes and examine them from different perspectives, in order to find an explanation for the unique effect of the side walls on the stability.

\begin{figure}
	\centerline{\includegraphics{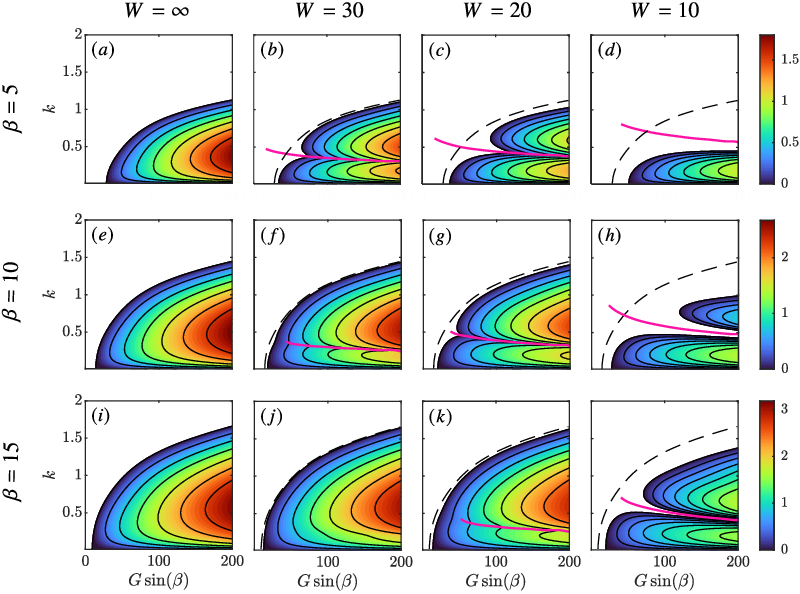}}
	\caption{Temporal growth rate contours in the $G\sin(\beta)-k $ space for (a)-(d)  $\beta=5$, (e)-(h) $\beta=10$, and (i)-(l) $\beta=15$, with $S=1$. The dashed line is the neutral curve of the $1D$ problem, while the purple line is a local minimum along the wavenumber.}
	\label{fig:contour1}
\end{figure}

\subsubsection{Initial observations: maximum temporal growth rate} \label{initial}
We examine the effect of spanwise confinement on the maximum temporal growth rate for a certain range of parameters. Figure \ref{fig:contour1} shows the temporal growth rate contours as a function of $G \sin(\beta)$ and $k$ for different $W$ and $\beta$ values. The white area is stable, while the colored area is unstable. The solid line is a local minimum in the growth rate along the wavenumber, while the dashed line is the neutral curve for the classical $1D$-mode. In general, we observe one unstable region identical to that of the $1D$ problem for $W = \infty$. As spanwise confinement becomes stronger, significant modifications are seen on both, the structure and the magnitude of the unstable region, while the local minimum in the growth rate shifts to higher wavenumbers. Moreover, the stabilisation effect shows to be  stronger for small inclination angles as seen when comparing the unstable regions of the inclination angles $\beta =5$ and $\beta=15$. 

A local minimum in the growth rate is shown at $W=30$ for inclination angle $\beta=10$. Decreasing the channel width to film height ratio to $W=20$, shifts the local minimum to higher wavenumbers, while the unstable region starts splitting. Further decrease in the channel width creates two unstable regions with a horizontal stable band in between. For a smaller inclination angle ($\beta=5$), the stabilisation effect is strong enough that the upper unstable region is completely stabilised, while the local minimum still exists. On the other hand, for $\beta=15$, the stabilisation effect is less dominant, however it manages to create the two unstable regions with a much thinner stable band separating them. {  Note that the mean film thickness ($h_m$) is kept constant while changing the channel width $W$, this is done by changing the flow rate $\dot{q}$ accordingly in order to keep the Reynolds number $Re$ constant, as equation (\ref{Reynolds}) shows.} \\
\begin{figure}
	\centerline{\includegraphics{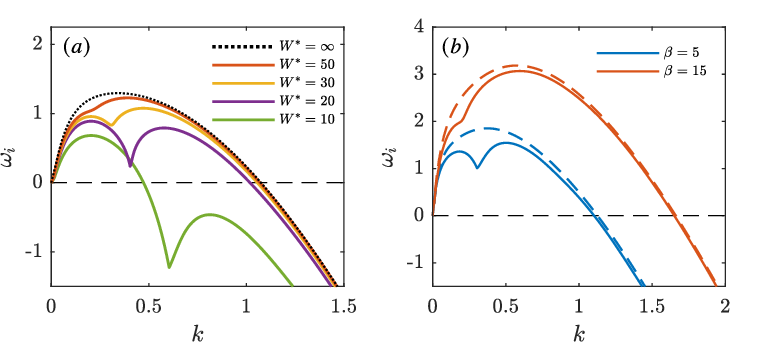}}
	\caption{Maximum growth rate for different $W$ values for $G\sin(\beta) =85$, $\beta = 10$, and $S=1$. (b) Influence of the inclination angle on the stabilisation effect for $W=30$, $G =500$, and $S=1$. The dashed lines correspond to the $1D$ growth rate.}
	\label{fig:omegavsk}
\end{figure}
A more detailed presentation of the stabilisation behaviour can be shown by plotting the maximum temporal growth rate against the wavenumber for a fixed $G\sin(\beta)$. In figure \ref{fig:omegavsk}(a), the growth rate at intermediate wavenumbers is affected the most, while it approaches the $1D$  growth rate at small and large wavenumbers. This agreement decreases as the confinement becomes stronger. A smooth dent in the growth rate curve is seen for moderate confinement ($W=50$), which slowly transforms into a sharp cusp as the channel becomes narrower. Furthermore, figure \ref{fig:omegavsk}(b) highlights the effect of the inclination angle on the stabilisation behavior. It is clear that the flow is stabilised more significantly at smaller inclination angles when comparing the difference between the growth rates for $W=30$ (solid) and that of the $1D$ problem (dashed). Also, the band of stabilised wavenumbers and the local minimum are located at higher values for smaller inclination angles.
To conclude this section, the fragmentation effect of the unstable region was first observed in the experimental results of \cite{kogel2020stability}, however no theoretical explanation was offered for this unique stabilisation behavior. We suggest an explanation for this phenomenon in the following sections using different perspectives of the problem.

\subsubsection{$2D$ stability modes} \label{twomodes}
The initial observations in the previous section were based on the maximum temporal growth rate $\omega_i$, regardless of what is the dominant stability mode. In the following, we examine several stability modes, and analyze the effect of confinement on them. The modes are obtained using the arc-continuation technique \citep{chan1982arc}. 
\begin{figure}
	\centerline{\includegraphics{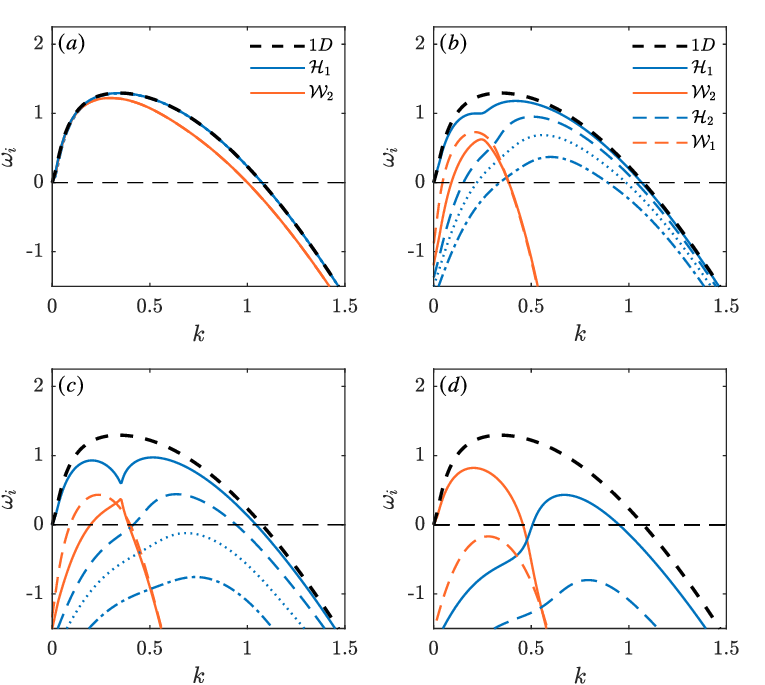}}
	\caption{Stability modes along the wavenumber for (a) $W=500$, (b) $W=40$, (c) $W=25$, (d) $W=15$ for the parameters $G=500$, $\beta=10$, and $S=1$. H-modes are blue, and W-modes are orange.   }
	\label{fig:fig5}
\end{figure}

Figure \ref{fig:fig5} shows the growth rate $\omega_i$ along the wavenumber $k$ for the most dominant stability modes for several degrees of lateral confinement. Starting with very weak confinement ($W = 500$) as in figure \ref{fig:fig5}(a), all the stability modes except for two, coincide all together matching the $1D$ mode (dashed line). Thus, these modes belong to the  {H-mode} type. The remaining pair of modes matches the  {H-modes} for small wavenumber, but diverges at large values. We will call these two modes wall modes (or simply  {W-modes}).

For $W = 40$, significant changes occur to all the stability modes as seen in figure \ref{fig:fig5}(b). All the modes are stabilised in general, but we observe three distinct stabilisation actions; the most unstable  {H-mode} is mainly stabilised at moderate wavenumbers showing the smooth indentation observed earlier in figure \ref{fig:omegavsk}(a). More interestingly, the remaining more stable  {H-modes} are strongly stabilised at low wavenumbers, while the opposite is experienced by the  {W-modes}, which show strong stabilisation at large wavenumbers.  Moreover, the local minimum of the most unstable  {H-mode} ($\mathcal{H}_1$) is accurately aligned with the maximum of the second most unstable  {W-mode} ($\mathcal{W}_2$). This configuration will oddly develop in later stages. {From another perspective, the H-modes (except for the most unstable mode) are transformed into waveguide modes. Waveguide modes are similar to the plane $1D$ mode, but allowed to propagate in the longitudinal direction only. They are also restricted to a cut-off wavenumber below which they do not propagate, which explains the stabilization of the H-modes for low wavenumbers.}

Figure \ref{fig:fig5}(c) shows similar stabilisation actions experienced by the  {H-modes} and  {W-modes} for $W=25$. Intriguingly, an attraction centre involving $\mathcal{H}_1$ and $\mathcal{W}_2$ modes takes a place leading to the formation of a sharp cusp in  $\mathcal{H}_1$ and an acute tip in  $\mathcal{W}_2$. The reason behind this attraction involving those two specific modes among the others will be investigated in subsequent sections. Eventually, further decrease in the channel width ($W = 15$) shows the final picture in figure \ref{fig:fig5}(d). The concurrence between the $\mathcal{H}_1$ cusp and $\mathcal{W}_2$ tip sufficiently increases until they meet and switch branches for wavenumbers before the meeting point. {At this stage the most unstable H-mode also becomes a waveguide mode}. This mode switching behaviour was also observed previously while studying the stability of falling liquid films on flexible substrates when changing the damping ratio \citep{alexander2020stability}. 

In summary, the solution of the biglobal stability problem contains several prominent stability modes. For the parameters under considerations, most of these modes belong to the {H-mode} type, while two of them are different, we name them  {W-modes}. The stabilisation effect at moderate wavenumbers is a result of two factors.  First, the spanwise confinement influence the two stability mode types in a different manner, {where H-modes are transformed into waveguide modes, and stabilized at low wavenumbers, while W-modes are stabilized at high wavenumbers.}  Second, an attraction centre involving two different modes, the most unstable  {H-mode} $(\mathcal{H}_1)$ and the second most unstable  {W-mode}$(\mathcal{W}_2)$. This attraction centre strengthens as confinement increases, until a mode switch takes place in the branch with lower wavenumbers, after which the most unstable H-mode becomes a waveguide mode as well. 

\begin{figure}
	\centerline{\includegraphics{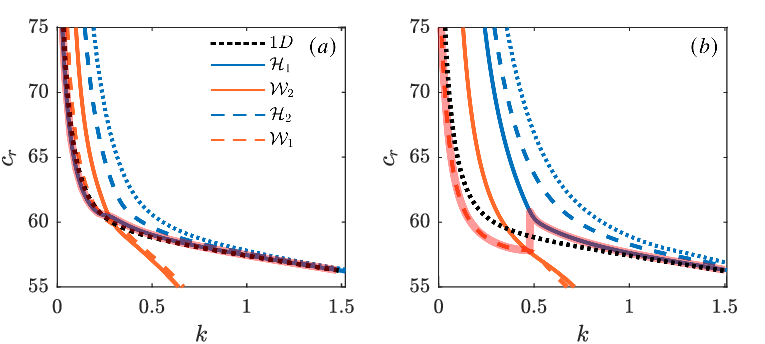}}
	\caption{{Phase velocity $c_r$ for different stability modes along the wavenumber for (a) $W=40$ and (b) $W=15$ for the parameters $G=500$, $\beta=10$, and $S=1$. The highlighted curve follows the maximum temporal growth rate along the wavenumber.}}
	\label{fig:cr_modes}
\end{figure}
{With the same approach, figure \ref{fig:cr_modes} shows the phase velocity ($c_r = \omega_r/k$) for the stability modes for $W=40$ and $W=15$. The highlighted curve follows the maximum temporal growth rate along the wavenumber. We observe a  similar behaviour to that of the temporal growth rate as a transition in the phase velocity is shown leading to a sharp jump when the channel width is sufficiently small. For $W=40$, $\mathcal{H}_1$ is still the most dominant stability mode along the wavenumber, and the phase velocity is very close to that of the $1D$ mode, while the other H-modes show higher phase velocity. The W-modes phase velocity drop sharply as wavenumber increases since they are stabilized strongly in this range.  As channel width decreases to $W=15$, the mode switch has occurred and a sudden jump in the phase velocity is observed. The phase velocity of the W-modes does not change significantly, but all the H-modes show a higher phase velocity than that of the $1D$ mode, which is a characteristic behaviour of waveguide modes \citep{elmore1985physics}. }

\subsubsection{Perturbation patterns and eigenvectors structure}
In this section, we study the perturbation patterns and the eigenvectors structure  of different stability modes for several confinement configurations. This will identify the structural nature of the different mode types, and help us better comprehend their particular behaviour as the channel width decreases. First, we compare the interface perturbation field $\tilde{h}(x,z)$ of the stability modes at two wavenumbers, i. e. $k=0.15$ and $k=0.75$. These two values are suitable options since the spanwise confinement affects the modes heavily at the former, but very lightly at the latter wavenumber. Afterwards, we plot the eigenmodes as a function of the wavenumber, along which the stability modes change significantly, in particular for strong confinement.

\begin{figure}
	\centerline{\includegraphics{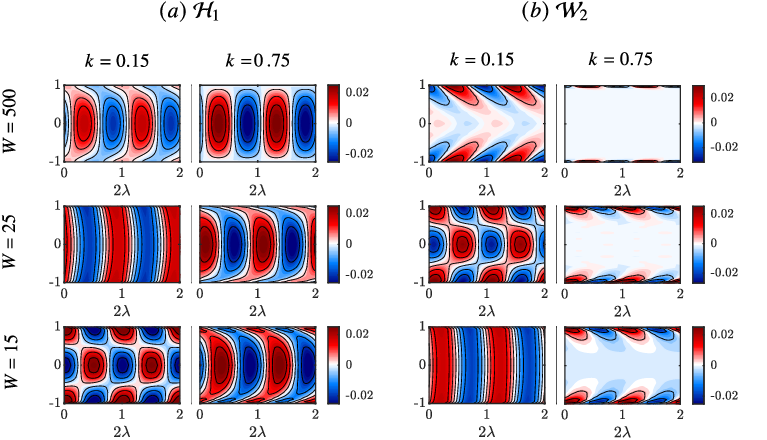}}
	\caption{Interface perturbation field $\tilde{h}(x,z)$ for the modes (a) $\mathcal{H}_1$ and (b) $\mathcal{W}_2$ at different wavenumbers for the parameters $G=500$, $\beta=10$, and $S=1$.}
	\label{fig:fig9}
\end{figure}

Figure \ref{fig:fig9} shows $\tilde{h}(x,z)$ for the pair of modes ($\mathcal{H}_1$,$\mathcal{W}_2$) for different $W$ values. The pair has an even symmetry around the axis $z=0$, which offers an explanation for the  unique behavior involving them. For $W=500$, the modes show their natural structure at $k=0.75$, where $\mathcal{H}_1$ is concentrated at the centre of the channel, while $\mathcal{W}_2$ is only observed in the vicinity of the side walls, as expected for a mode created by the confinement. At $k=0.15$, the two modes are in close proximity, see figure \ref{fig:fig5}(a)), and thus they affect each other slightly causing $\mathcal{H}_1$ to increase its curvature, while $\mathcal{W}_2$ spreads more into the domain.  These structures are logical since confinement is yet very weak. For  $W=25$, $\mathcal{H}_1$ changes its structure to a straight wave along the spanwise direction at $k=0.15$, while it keeps the original structure at $k=0.75$ with more curvature. For the same confinement ($W=25$), $\mathcal{W}_1$ is dramatically enhanced at $k=0.15$, while it just spreads slightly into the domain at $k=0.75$. Finally, $W=15$ shows a structure switch between the two modes at $k=0.15$. It also shows a strong curvature in $\mathcal{H}_1$ at $k=0.75$, while stronger spreading for $\mathcal{W}_1$.

\begin{figure} 
	\centerline{\includegraphics{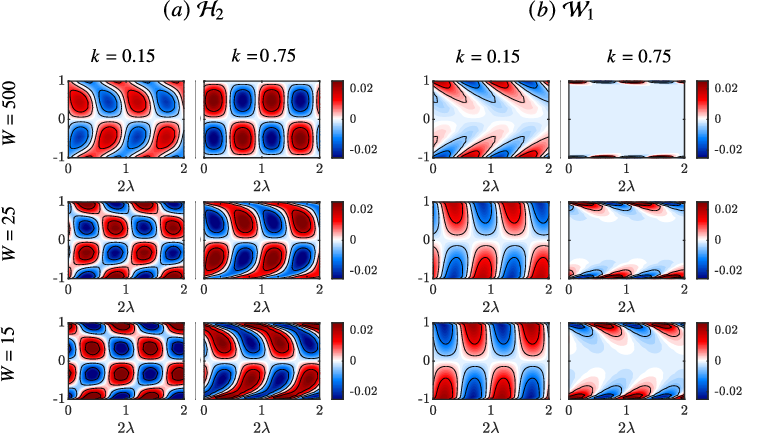}}
	\caption{Interface perturbation field $\tilde{h}(x,z)$ for the modes (a) $\mathcal{H}_2$ and (b) $\mathcal{W}_1$ at different wavenumbers for the parameters $G=500$, $\beta=10$, and $S=1$.}
	\label{fig:fig10}
\end{figure}

For a complete analysis, we also plot the interface perturbation field $\tilde{h}(x,z)$ for the modes pair ($\mathcal{H}_2$,$\mathcal{W}_1$), which are the second most unstable  {H-mode} and the most unstable  {W-mode}, respectively (Figure \ref{fig:fig10}). These two modes have an odd symmetry, such as $\tilde{h}(x,z)=-\tilde{h}(x,-z)$, and show less complex behaviour compared to the other pair ($\mathcal{H}_1$,$\mathcal{W}_2$). For $W=500$, the spatial structure  $\mathcal{H}_2$ is one step higher in the harmonic range compared to $\mathcal{H}_1$, while $\mathcal{W}_1$ is restricted at the walls. As confinement becomes stronger ($W=25$,$W=15$),  $\mathcal{H}_2$ changes structure and shows higher harmonics of $\mathcal{W}_1$ at $k=0.15$, while it keeps its original structure with a curvature proportional to the confinement degree at $k=0.75$.  Furthermore, $\mathcal{W}_1$ spreads more into the domain for all wavenumbers, but strongly at lower values. For both mode pairs, the symmetry of the perturbation fields are conserved for any $W$ value.

\begin{figure}
	\centerline{\includegraphics{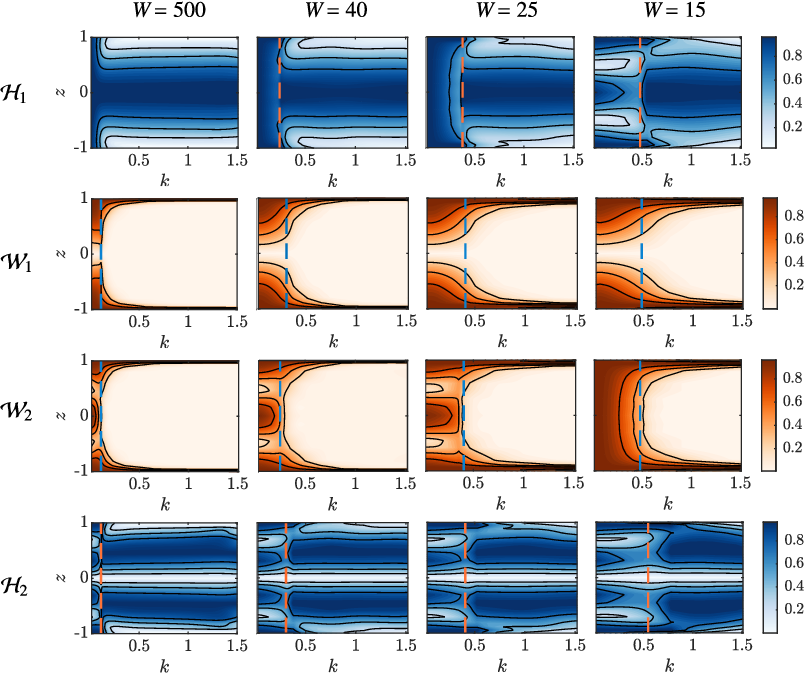}}
	\caption{ The real part of normalized $\eta(z)$ eigenmode along the wavenumber for different confinement ratios, for the parameters $G=500$, $\beta=10$, and $S=1$.}
	\label{fig:fig7}
\end{figure}
With regards to the eigenmodes,  figure \ref{fig:fig7} shows the absolute of the normalized $\eta(z)$, which is the eigenmode of the interface perturbation field $\tilde{h}(x,z)$ (see equation \eqref{normal_modes}). The first row  belongs to the H-mode $\mathcal{H}_1$, while the second row belongs to most unstable  {W-mode} ($\mathcal{W}_1$). The third and fourth rows show the eigenmodes $\mathcal{W}_2$  and $\mathcal{H}_2$, respectively. The dashed lines present the location at which the different mode types cross each other, except for $\mathcal{H}_1$ at $W=40$ and $W=25$ where it shows the local minimum.   When the spanwise confinement is very weak ($W=500$), all the eigenmodes are constant along the wavenumber except at small values. The most dominant mode $\mathcal{H}_1$ is concentrated at the center and zero at the walls, while $\mathcal{H}_2$ has similar structure but with a higher step in the harmonic range. On the other hand, the  {W-modes} are strictly concentrated at the side walls and zero everywhere else. The stability modes form a "discrete spectrum" consisting of the  {H-modes} only ($\mathcal{H}_1$,$\mathcal{H}_2$,$\mathcal{H}_3$, etc.), while  {W-modes} play no major role. This is in agreement to what is observed earlier in figures \ref{fig:fig9} and \ref{fig:fig10}.

The unique patterns in the eigenmodes at small wavenumbers for $W=500$ spread into higher values as confinement increases to $W=40$ and $W=25$. The presence of the  {W-modes} become more evident causing significant restructuring of the other eigenmodes along the wavenumber axis. The region in which the eigenmodes are reconstructed extends into higher wavenumbers proportionally to the strength of the confinement. In this limited region (bounded by the dashed line), the stability modes form a different harmonic sequence; $\mathcal{W}_1$ is no longer  only concentrated at the walls but spreads towards the center of the channel, while $\mathcal{W}_2$ shows similar structure but with the further step in the harmonic range.  Moreover, the  {H-modes} are reconstructed to fit the new harmonic sequence; $\mathcal{H}_2$ shows a third step in the new harmonic range, while $\mathcal{H}_1$ shows a constant maximum along $z$, which is the lowest step in the harmonic range. The eigenmodes are slightly affected after the cusp (dashed line), and the original harmonic series is conserved. For $W=15$, the same structures spread to higher wavenumbers and become more prominent. Also, we notice the switch of the eigenmode between $\mathcal{H}_1$ and $\mathcal{W}_2$ in order to conserve the harmonic series.

In short, when the spanwise confinement is negligible, the  {H-modes} form a harmonic sequence with a fundamental mode that is maximum at the center and zero at the side walls, while  {W-modes} are insignificant. When confinement becomes stronger, the  {W-modes} presents an opposite (as in maximum at the walls and zero in the center of the channel) harmonic series at small wavenumbers region. This forces a reconstruction on the  {H-modes} to be compatible in the new harmonic sequence.

\subsubsection{Eigenvalue spectra of confined 2D problem vs. 1D oblique problem}
As seen, the W-modes play no significant role for weak spanwise confinement, but experience dramatic change in both growth rate  and spatial structure for strong confinement. We also observed that they seemed to developed from H-modes when confinement is weak. In this section, we confront the eigenvalues spectrum of the $2D$ problem to that of the oblique unbounded case to show how the two W-modes evolve as confinement becomes stronger. In the oblique simplified problem, the base state is spanwise invariant ($\bar{U}(y),\bar{P}(y)$), while the perturbation fields have a harmonic structure in the spanwise direction, and expanded as follows:

\begin{subeqnarray}
	&  	\tilde{\boldsymbol{q}}(x,y,z,t) = \boldsymbol{\hat{Q}}(y) \mbox{ exp}(i \zeta z) \mbox{ exp}(ikx - i\omega t),\\
	& \tilde{h}(x,z,t) = \hat{\eta} \mbox{ exp}(i \zeta z) \mbox{exp}(ikx - i\omega t),
	\label{oblique}
\end{subeqnarray}

\begin{figure}
	\centerline{\includegraphics{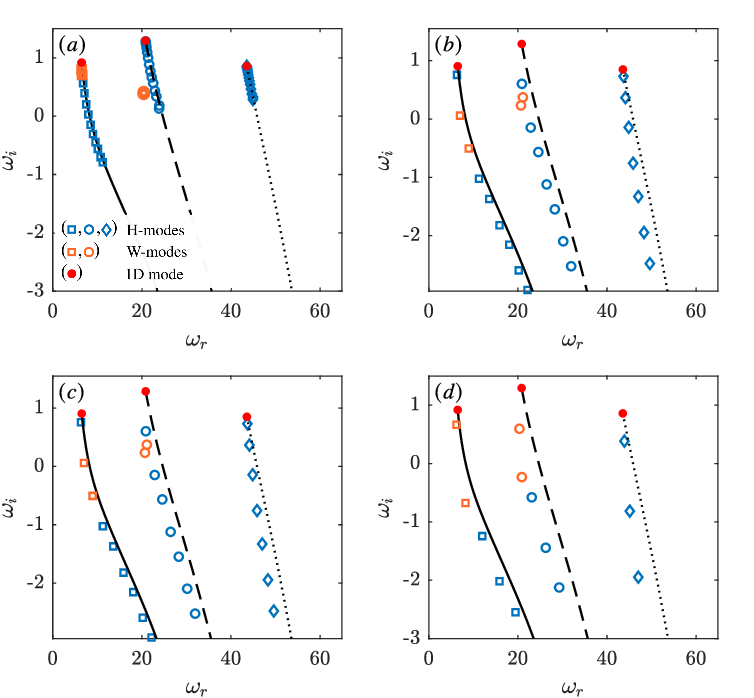}}
	\caption{ $2D$ problem eigenvalues spectrum (hollow markers) vs. oblique dispersion relation (lines) as a function of the spanwise wavenumber $\zeta$. The red dots mark the $1D$ mode, while the blue and orange markers are the $2D$ H-modes and W-modes, respectively. (a) $W=100$, (b) $W=40$, (c) $W=25$, (d) $W=15$ at wavenumber values $k=0.15$ ($\square$,	solid), $k=0.35$ ($\circ$, dashed), $k=0.75$ ($\diamond$, dotted). }
	\label{fig:fig12}
\end{figure}

where $\zeta \in \mathbb{R}$ is the spanwise wavenumber. Equation \eqref{oblique} implies that the spanwise direction is unbounded, since any spanwise wavelength is possible for the perturbations ($\zeta \in \mathbb{R}$).  For $\zeta=0$, the oblique problem becomes identical to the $1D$ problem, while for $\zeta \ne 0$, a spanwise variation is allowed for the perturbations without considering the confining effect of the side walls. Similarly to the $1D$ problem, only one unstable mode (H-mode) shows in the oblique problem regardless of the value of $\zeta$.

 {Figure \ref{fig:fig12} compares the eigenvalues spectrum of the $2D$ problem to the oblique mode. The two problems are compared at three different wavenumbers for several confinement values. The oblique solutions in all the subfigures are the same since the problem is not a function of the spanwise confinement. For $\zeta=0$, the oblique mode (line) matches the $1D$ mode. As $\zeta$ increases, the oblique modes become more stable since they experience stronger viscous damping. This is a consequence of the increase in the diffusive term weight as the Laplacian term became ($-k^2+\partial_{yy}-\zeta^2$) instead of ($-k^2+\partial_{yy}+ \partial_{zz}$) due to the oblique expansion in equation \eqref{oblique}.}

Focusing first on $W=100$ (figure \ref{fig:fig12}(a)), the $2D$ modes including both  {H-modes} and W-modes match the oblique dispersion relation at small wavenumber $k=0.15$, while being tightly packed around the $1D$ mode. For larger wavenumbers, the H-modes follow the oblique mode but the W-modes start departing from the line ($k=0.35$), until they disappear at $k=0.75$.  For stronger  confinement as in figure \ref{fig:fig12}(b-d), the $2D$ modes spread out on the oblique line, while the agreement decreases. The most evident discrepancy between the $2D$ modes and the oblique dispersion relation is observed for moderate and high wavenumbers. This discrepancy is a reason of the complex behavior  we observed involving the stability modes, which takes place at moderate wavenumbers and influence higher wavenumbers as well. Interestingly, the $2D$ modes including the W-modes are very close with the oblique dispersion relation for strong confinement at small wavenumber $k=0.15$. Moreover, the $2D$ modes are always more stable than the $1D$ mode for all parameters.
 
\subsection{Discussion on surface tension and contact line dynamics} \label{contact}
{The results presented are limited to cases where  surface tension is not the dominant restoring force where $S \sim \mathcal{O} (1)$ in both theoretical analysis and experiments. In such scenarios, the capillary length is very small compared to the channel width, for example Bo is of $ \mathcal{O}(10^3)$ in experiments. Therefore, neglecting the wetting effects and using the simple free-end contact line boundary condition (2.3) is admissible. This is also consistent with the base flow under consideration in which the interface is assumed to be invariant in the spanwise direction. }

{Nevertheless, capillary dissipation due to wetting effects and contact line dynamics in confined surface waves is non-negligible in several cases, such as, very narrow channels \citep{monsalve2022space}, micro-gravity conditions \citep{zhang2013capillary}, or simply when surface tension is the main restoring force, and capillary forces are dominant over viscous forces \citep{georgantaki2011measurements}. The last scenario is directly related to this work where it was shown experimentally that surface tension, when dominant, can significantly affect the  onset of the longwave instability in falling films confined by side walls. They suggest that this effect is a result of a 3D capillary wave attenuation mechanism, which stems from damping the oscillations near the side walls caused by (i) viscous boundary layer, and more relevantly, (ii) the heightened resistance to the depinning of the contact line.}

{Under the stability model in this work,  the free-end boundary condition (2.3) allows the interface to freely slip at the side walls, which eliminates the damping effect due to resistance of contact line depinning. Therefore, the unique capillary damping of oscillations near the side walls, which leads to a delay in the long-wave instability onset, is not captured in our model. In order to be able to analyse scenarios in which surface tension is dominant, a more complex contact line boundary condition is necessary. \cite{hocking1987damping}, motivated by experimental evidence suggesting that contact angle is not fixed, but depends on contact line motion \citep{dussan1979spreading}, proposed a contact line boundary condition in which the contact line velocity is proportional to the variation in the contact angle $\theta$:} 
\begin{equation}
\frac{\partial h}{\partial t}  = \alpha \frac{\partial h}{\partial z},
\end{equation}\label{hocking}
{ where $\alpha$ is a slip coefficient that characterizes the boundary condition. An important feature of this boundary condition is that it includes two extreme limits of zero and infinite slip coefficients, which correspond to a pinned-end ($\partial_t h =0$) and a free-end ($\partial_z h =0$) conditions, respectively. More complex models are present in the literature that incorporate contact angle hysteresis, where $\alpha$ is described as function of contact line displacement \citep{viola2018capillary}.}

{Another related issue to be addressed here is the well-known contact line singularity, which results from assuming a moving contact line at the wall alongside the no-slip boundary condition \citep{davis1974motion}. This singularity can be avoided by using a pinned-end contact line boundary condition which is physically relevant in specific cases, such as brim-full containers \citep{benjamin1979gravity}. Apart from that, various remedies were developed to remove the singularity, such as, the common approach of using a slip model in which the fluid is assumed to slip in a small neighborhood near the contact line \citep{dussan1979spreading}. The outer region where the no-slip applies, and the slip region can be perfectly matched as done in \cite{sibley2015asymptotics}. For a comprehensive comparison between approaches to tackle this singularity, please refer to \cite{snoeijer2013moving,sibley2015comparison}.  } 
 
{In this context, using the range of parameters relevant to this work, we compared the results produced by both limits of Hoking's Law \eqref{hocking}, a pinned contact line versus a free-moving contact line. We found the difference in temporal growth rate to be clearly visible but consistently below $5\%$. This shows that for the range of parameters we considered, the implementation of a free-moving contact line condition and the associated singularity do not affect the qualitative accuracy of our results. Further investigation of capillary-dominated flows is planned for future work.}




\section{Conclusions} \label{sec:conclusion}
Using temporal stability analysis and experimental approaches, we study the linear stability of falling liquid films bounded by side walls. The two  agree very well for small and large wavenumbers. Noticeable discrepancies were observed before and after the cusp in the growth rate. We showed that the supremum of the resolvent norm is large in these two regions, indicating the presence of stronger receptivity to disturbances. We plan to investigate this further in a future work. 

The theoretical stability model showed that the solution of the confined $2D$ problem shows several   stability modes, unlike the classical $1D$ problem where only the  {H-mode} is present. In the $k-G \sin(\beta)$ plane, we observe the fragmentation effect on the unstable region which was reported by \cite{kogel2020stability}. The unstable region splits into two smaller regions separated by a stable band at moderate wavenumbers. As confinement becomes stronger, the upper unstable region disappears. The stabilising effect is depending on the inclination angle. As  inclination angle decreases, the magnitude of damping increases, and also the band of stabilised wavenumbers shifts towards higher values. 

The reason behind this behavior is the existence of different stability modes. For a certain range of parameters,   the $2D$ modes except a pair converge to the $1D$ {H-mode} for weak confinement configurations. The unique pair is named {W-modes}. As confinement strengthens, {H-modes} and {W-modes} are stabilised in a different manner along the wavenumber. The former is stabilised strongly at low wavenumbers, while the latter experience stabilisation at high wavenumbers. An attraction centre involving the most unstable {H-mode} and a more stable {W-mode} also contributes to the formation of the stabilised wave band. The spatial structure of the eigenmodes shows an alternative even-odd symmetry in both {H-modes} and {W-modes}. The spatial structure of the most unstable {H-mode} shows a maximum in the center of the channel, while that of the most unstable {W-mode} shows a maximum at the walls. The less stable modes are an alternating symmetry with a further degree in the harmonics. The eigenvalues spectra shows that the $2D$ eigenvalues follow the one-dimensional oblique problem for small wavenumbers regardless of the confinement degree, and it deviates at larger wavenumbers as confinement becomes stronger.

\section{Future work}
 {The results presented in this work can stimulate  future research.  The biglobal stability problem results in a multiple stability modes with comparable growth rates. This could lead to very rich non-linear dynamics, especially that the stability modes are structural harmonics of each other \citep{ducimetiere2021effects}. One way to examine this can be through three-dimensional direct numerical simulations which was previously carried out for falling films but with periodic  boundary conditions in the spanwise direction \citep{dietze2014three}. Moreover, the excitation introduced by the paddle can be included in the theoretical model as a forcing term. The forced problem can also be solved and compared to the existing autonomous problem as in the work of \cite{fabre2020acoustic}.  Finally, there exist a plethora of applications that involve volatile liquid films falling down a heated surface. The instabilities induced by thermocapillary forces \textit{(S-mode}) or vapour recoil (\textit{E-mode}) were only examined in the two dimensional configurations in the classical Orr-Sommerfeld problem \citep{mohamed2020linear} or in the long wave expansion regime \citep{mohamed2021spatiotemporal}. The analysis of the effect of spanwise confinement on this kind of instabilities is of great benefit due to the intrinsic spanwise nature of the thermal instabilities. }

\section*{Acknowledgements} 
 The authors would like to acknowledge Marion Märkl for helping in carrying out the experiments.{ The authors would like to thank the reviewers for their insightful comments and suggestions, which helped improve the quality of this work.}\\

{\bf \noindent Declaration of Interests.\\}
The authors report no conflict of interest.

\bibliographystyle{jfm}
\bibliography{jfm}

\end{document}